\documentclass[prl,reprint,a4paper,superscriptaddress,showpacs,citeautoscript,nofootinbib]{revtex4-1}

\pdfoutput=1
\usepackage[charter]{mathdesign}
\usepackage{amsmath}

\usepackage[pdftex]{graphicx}
\usepackage{microtype}
\usepackage{bm}\let\vec\bm

\usepackage[svgnames]{xcolor}
\usepackage[
	colorlinks=True,linkcolor=DarkRed,citecolor=ForestGreen,urlcolor=MediumBlue,
	pdfstartview=FitH,bookmarks=False,pdfpagemode=UseNone
]{hyperref}

\newcommand{\Li}{_{\mathrm{Li}}}
\newcommand{\OLi}{_{0\mathrm{Li}}}
\newcommand{\K}{_{\mathrm{K}}}
\newcommand{\OK}{_{0\mathrm{K}}}

\begin{document}

\title{\boldmath Interplay of the pseudogap and the BCS gap for heteropairs in $^{40}$K--$^6$Li mixture}

\author{P. Lipavsk\'y}
\affiliation{Faculty of Mathematics and Physics, Charles University, Ke Karlovu
3, 12116 Prague 2, Czech Republic}
\author{C. Berthod}
\affiliation{Department of Quantum Matter Physics, University of Geneva, 24 quai
Ernest-Ansermet, 1211 Geneva, Switzerland}

\date{September 1, 2016}

\begin{abstract}

The description of heteropairs like $^{40}$K--$^6$Li near and in the
superconducting state requires a fully selfconsistent theory [see
\href{http://journals.aps.org/pra/abstract/10.1103/PhysRevA.90.043622}{Hanai and
Ohashi, Phys. Rev. A \textbf{90}, 043622 (2014)}]. We derive analytic pseudogap
Green's functions for the `normal' and superconducting states from the
Luttinger-Ward theory with the T-matrix in the static separable approximation.
The selfconsistency in the closing loop of selfenergy has two pronounced effects
on the single-particle spectrum. First, the single-particle excitations decay
before the asymptotic quasiparticle propagation is established, therefore the
normal state is not a Fermi liquid. Second, the pseudogap has a V shape even for
s-wave pairing. The V-shaped pseudogap and the U-shaped BCS gap interfere
resulting in slope breaks of the gap walls and the in-gap states in the density
of states. Various consequences of an incomplete selfconsistency are
demonstrated.

\end{abstract}

\maketitle

The cold Fermi gas of 1:1 mixture of $^{40}$K and $^6$Li  is expected to develop
an unconventional superfluidity with Cooper pairs composed of different species
\cite{PSS04}. If achieved, this phase transition offers experimental
verification of theories of heteropairs condensation important also in other
fields of physics, e.g., the color superconductivity \cite{ASRS08}. Recently
Hanai and Ohashi \cite{HO14} have discussed the critical temperature using the
Thouless criterion based on the instability of the normal state. They found that
the two-particle T-matrix theory has to be fully selfconsistent to describe the
superconducting phase transition. Failure of partly or fully non-selfconsistent
approaches can be traced down to their ill behaved density equation of state
\cite{MFDVLH15}.

The fully selfconsistent T-matrix approximation known as the Luttinger-Ward (LW)
theory \cite{BPE14} results in a qualitatively correct density equation of state
\cite{AD15}, but it suffers from two setbacks when applied to the pseudogap and
superconducting states. First, there is no analytic formula for the pseudogap
density of states (DOS) so that each observed signature requires independent
numerical study. Second, although the T-matrix includes the pairing channel, the
LW theory fails to describe the Bardeen, Cooper and Schrieffer (BCS) gap and it
has to be modified to cover the superconducting state. This extension is not
unique because several modifications are possible, which lead to distinguishable
predictions for the DOS.

In this paper we derive an approximation convenient for eventual fits to
experimental data on the BCS side of the BCS-BEC crossover. We start from
analyzes of single-particle excitations in the LW theory of the pseudogap state.
Assuming the dominant role of low-lying preformed pairs, we adopt the static and
separable approximation of the T-matrix. In this approximation we obtain an
analytical Green's function from which it is shown that the lifetime of
excitations is the same as the time scale of binary correlations, therefore the
quasiparticle concept is not applicable and the system is not a Fermi liquid.
The corresponding DOS has the pseudogap. In spite of the s-wave pairing, the
pseudogap has a V shape which compares well with numerical results of Hanai and
Ohashi \cite{HO14}. Then we modify the theory for the superconducting state and
derive an analytic Green's function which covers the combined effect of the
V-shaped pseudogap and the U-shaped BCS gap on the single-particle excitation
spectrum. We discuss approximations with and without feedback effect of the BCS
gap on the pseudogap, which lead to substantially different resulting DOS.

Although we discuss a rather general effect of the selfconsistent Green's
function in the closing loop on the single-particle energy spectrum, we prefer
to relate all formulas to the specific example of $^{40}$K-$^6$Li gas and use
the $^{40}$K and $^6$Li atomic masses in plots of the spectral functions.

The bare Green's function for lithium atoms
$G^R\OLi(k)=1/\left(\omega+i0-\epsilon\Li\right)$ depends on the kinetic energy
$\epsilon\Li=|\vec{k}|^2/(2m\Li)-\mu\Li$ counted from the chemical potential.
The only important interaction is between lithium and potassium atoms. It is
tuned by the magnetic field with the Feshbach resonance enhancement in the
scattering s-channel \cite{PSS04}. The selfconsistent pseudogap-state Green's
function $G^R\Li(k)=G^R\OLi(k)+G^R\OLi(k)\Sigma^R\Li(k)G^R\Li(k)$ depends on the
Li-K interaction via the selfenergy
	\begin{multline}\label{selfenergy1}
		\Sigma^R\Li(k)=\int\frac{d\Omega}{2\pi}\frac{d\vec{Q}}{(2\pi)^3}
		T^R(Q)G^<\K(Q-k)\\
		-\int\frac{d\Omega}{2\pi}\frac{d\vec{Q}}{(2\pi)^3}T^<(Q)G^A\K(Q-k),
	\end{multline}
where $k\equiv(\omega,\vec{k})$ and $Q\equiv(\Omega,\vec{Q})$. We assume
non-retarded contact interaction so that the T-matrix depends only on the
four-momentum of the interacting pair. In the Fourier picture the contact
interaction is independent of momentum, therefore the wave function of pairs is
isotropic having the s-wave symmetry.

The functions $G^<\K(Q-k)=2\mathrm{Im}G^A\K(Q-k)f_{\mathrm{FD}}(\Omega-\omega)$
and $T^<(Q)=-2\mathrm{Im}T^R(Q)f_{\mathrm{BE}}(\Omega)$ include statistics. The
potassium function remains small as $f_{\mathrm{FD}}<1$ while the two-particle
function is large at very small frequencies $\Omega$ that appear only for very
small pair momenta $\vec{Q}$. Based on the dominant role of pairs with small
four-momentum we adopt the approximation
	\begin{equation}\label{selfenergy2}
		\Sigma^R\Li(k)\approx-D^2G^A\K(-k),
	\end{equation}
with $D^2=(2\pi)^{-4}\int d\Omega d\vec{Q}\,T^<(Q)$. In the following we take
the value of $D$ as a parameter and focus on properties of the single-particle
spectral function and the density of states.

Equation~\eqref{selfenergy2} recalls the approximation of the pseudogap
introduced by Maly \textit{et al.}\ \cite{MJL99} who have numerically verified
the dominant role of the low energy and momentum part of the T-matrix. In spite
of this common point, our results sharply differ from \cite{MJL99}, because we
start from the LW theory so that the fully selfconsistent Green's function
closes the loop of the selfenergy \eqref{selfenergy2}, while studies of the
pseudogap by Kathy Levin's group \cite{MJL99,CSTL05} are based on the partly
selfconsistent and partly non-selfconsistent theory of Kadanoff and Martin (KM)
\cite{KM61} with the bare Green's function in the closing loop.

\begin{figure}[tb]
\centering{\includegraphics[width=0.8\columnwidth]{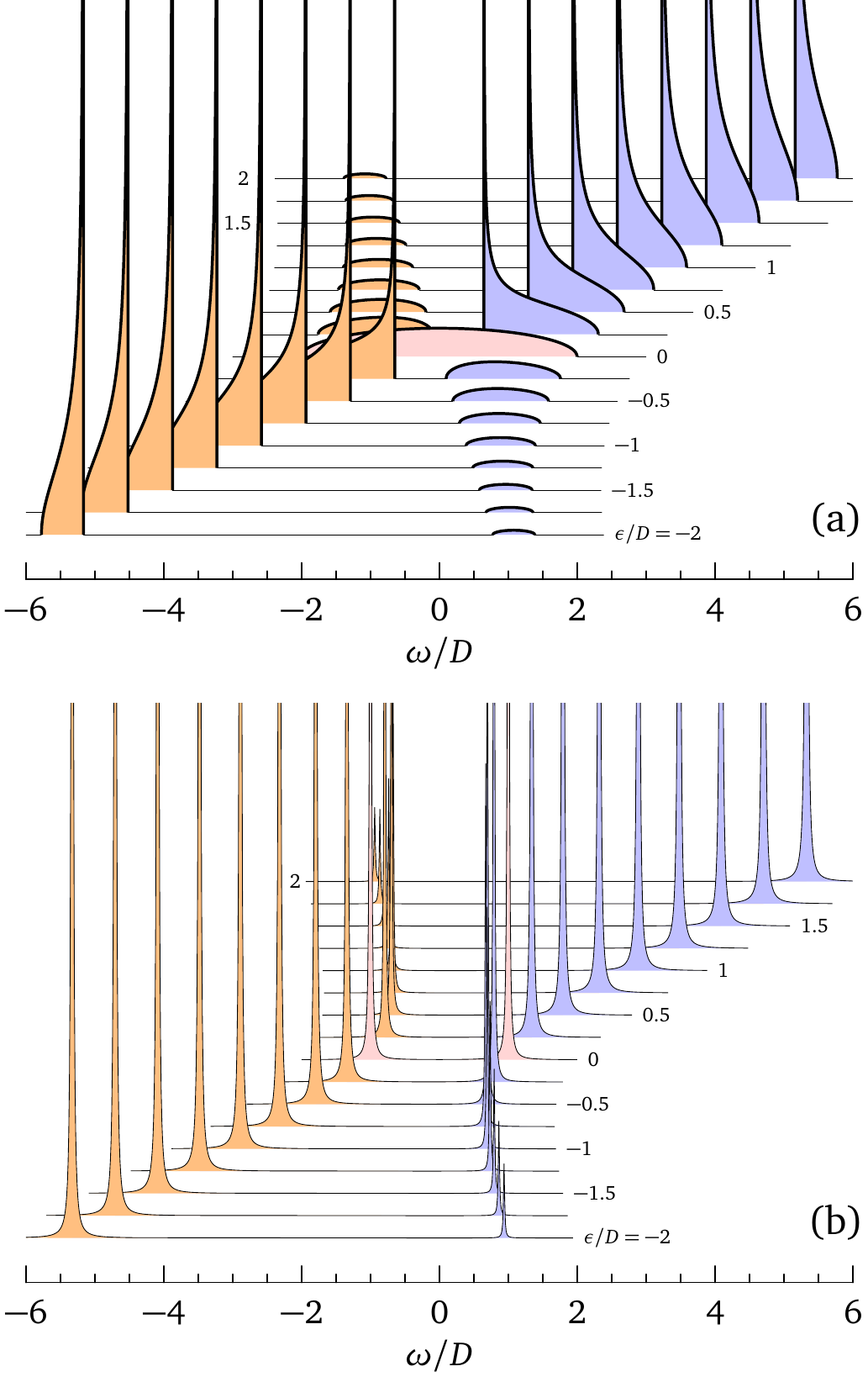}}
\caption{\label{fig1}
Spectral function $A\Li(k)$ in the pseudogap state (a) with the selfconsistently
closed loop and (b) non-selfconsistently closed loop, for
$\epsilon\Li=\epsilon\sqrt{m\K/m\Li}$, $\epsilon\K=\epsilon\sqrt{m\Li/m\K}$, and
different values of $\epsilon$. The curves are shifted vertically for clarity. A
Lorentzian broadening $10^{-2}D$ was applied in (b) to represent
$\delta$-function peaks. Due to higher potassium mass the back-folded branches
are flatter, which gives the illusion of a 3D perspective.
}
\end{figure}

The advanced Green's function of potassium satisfies an analogous equation with
$\Sigma^A\K(-k)\approx-D^2G^R\Li(k)$. The set of these two equations has solution
	\begin{equation}\label{Gpseudogap}
		G^R\Li(k)=\frac{\omega+\epsilon\K}{2D^2}\left(1-\eta
		\sqrt{\left|1-\frac{4D^2}{(\omega+\epsilon\K)(\omega-\epsilon\Li)}
		\right|}\,\right),
	\end{equation}
with $\eta=i\,\mathrm{sign}(2\omega+\epsilon\K-\epsilon\Li)$ if
$0<(\omega+\epsilon\K)(\omega-\epsilon\Li)<4D^2$ and $\eta=1$ otherwise. The
spectral function $A\Li(k)=(-1/\pi)\mathrm{Im}G^R\Li(k)$ is shown in
Fig.~\ref{fig1}(a). It is non-zero in two energy intervals
$\frac12(\epsilon\Li-\epsilon\K)\!-\!\sqrt{
\frac14\left(\epsilon\Li+\epsilon\K\right)^2+4D^2}
<\omega<{\min}(-\epsilon\K,\epsilon\Li)$, $\max(-\epsilon\K,\epsilon\Li)<\omega<
\frac12(\epsilon\Li-\epsilon\K)+
\sqrt{\frac14\left(\epsilon\Li+\epsilon\K\right)^2+4D^2}$.

\begin{figure}[tb]
\centering{\includegraphics[width=0.8\columnwidth]{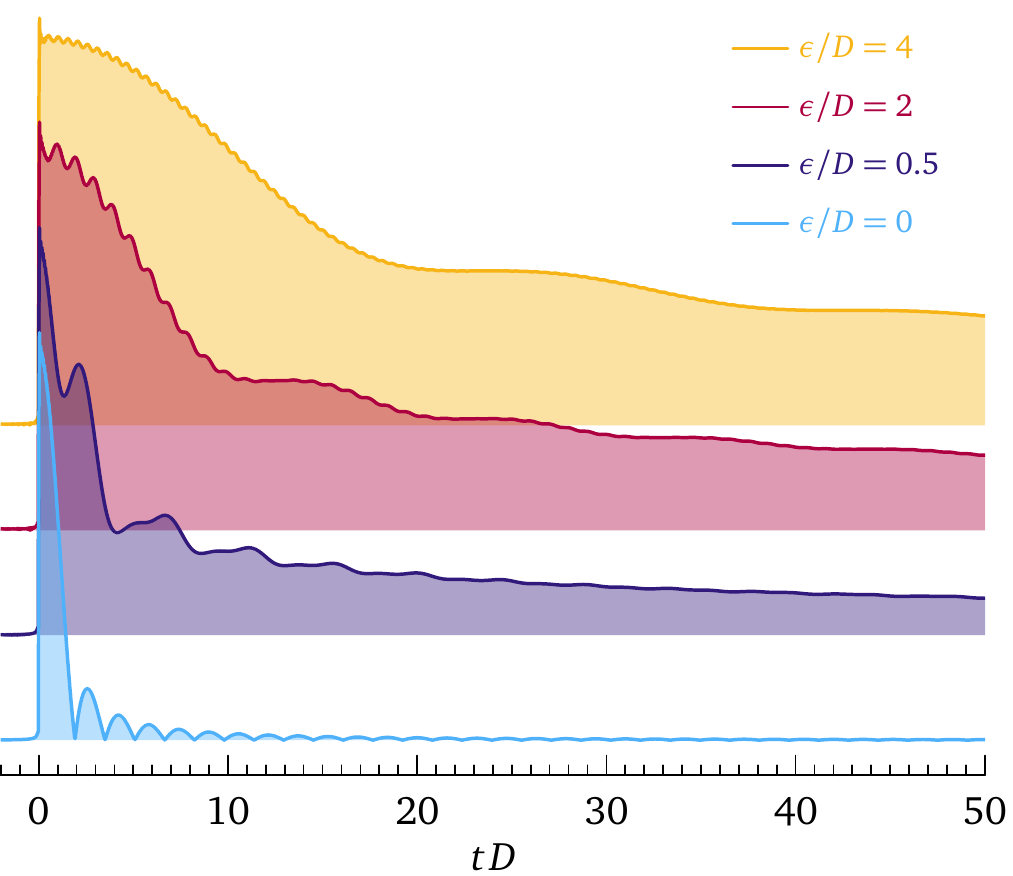}}
\caption{\label{fig2}
Time-dependent excitations in the pseudogap state: $\big|G^R\Li(t,\vec{k})\big|=
\big|(1/2\pi)\int_{-\infty}^{\infty}d\omega\,G^R\Li(\omega,\vec{k})\exp(-i\omega
t)\big|$ is shown for $\epsilon\Li=\epsilon\sqrt{m\K/m\Li}$,
$\epsilon\K=\epsilon\sqrt{m\Li/m\K}$, and different values of $\epsilon$.
}
\end{figure}

For comparison we show in Fig.~\ref{fig1}(b) the spectral function of the KM
theory, which obtains using $\Sigma^R\Li(k)=-D^2G^A\OK(-k)$ instead of the
selfenergy \eqref{selfenergy2}. The KM spectral function is of the BCS type
having two $\delta$-functions at energies
$\omega=\frac12(\epsilon\Li-\epsilon\K)\pm
\sqrt{\frac14(\epsilon\Li+\epsilon\K)^2+D^2}$. Apparently, with the same
approximation of the T-matrix, the KM and LW theories predict very different
single-particle excitations. The KM theory results in quasiparticles of infinite
lifetime, while the LW theory leads to excitations with non-exponential decay,
see Fig.~\ref{fig2}. Both spectra show back-folding near the Fermi level which
is experimentally confirmed for balanced gases
\cite{GSDJPPS10,PPPSSGDJ11,FFVKK11}. The spectra differ, however, in the line
shape. The $\delta$-peak of the KM theory broadened by the lifetime is symmetric
while the LW `line' has higher weight on the side of Fermi level in agreement
with the momentum-dependent spectroscopy \cite{PPPSSGDJ11}. Another trend which
seems to be consistent with \cite{PPPSSGDJ11} is that the line becomes broader
on approaching the Fermi surface; in a Fermi liquid it would be the opposite.

From the selfenergy \eqref{selfenergy2} one can see that the time decay of the
potassium propagator is the same as the time decay of the lithium selfenergy.
Since the time scales of propagators and selfenergies are comparable, the
quasiparticle picture is not applicable. An attempt to define the quasiparticle
energy from $1/G^R\Li(\varepsilon\Li,\vec{k})=0$ yields the bare particle of
infinite lifetime $\varepsilon\Li=\epsilon\Li$ and the weight of the pole is not
defined because $1/(1-\partial{\rm Re}\Sigma\Li/\partial\omega)$ has
discontinuity at energy $\epsilon\Li$. Indeed, from Eq.~\eqref{Gpseudogap} one
finds $G^R\Li\propto 1/\sqrt{\omega-\epsilon\Li}$ which cannot be approximated
by the quasiparticle relation $\propto 1/(\omega-\epsilon\Li)$. Clearly, the
pseudogap state is not a Fermi liquid.

To enlighten consequences of the mass and density imbalance, we express the
imbalanced function \eqref{Gpseudogap} in terms of the balanced one by the
substitutions $\epsilon\Li=\epsilon'\alpha-\mu\Li$ and
$\epsilon\K=\epsilon'/\alpha-\mu\K$ with $\alpha=\sqrt{m\K/m\Li}$,
$\epsilon'=|\vec{k}|^2/(2\sqrt{m\K m\Li})$,
$\varepsilon=\epsilon'+\omega(\alpha-1/\alpha)/2-(\mu\K\alpha+\mu\Li/\alpha)/2$,
and $w=\omega(\alpha+1/\alpha)/2-(\mu\K\alpha-\mu\Li/\alpha)/2$, as
	\begin{equation}\label{GpseudogapBal}
		G^R\Li(k)=\frac{1}{\alpha}\frac{w+\varepsilon}{2D^2}\left(1-
		\eta\sqrt{\left|1-\frac{4D^2}{w^2-\varepsilon^2}\right|}\,\right).
	\end{equation}
$\eta=i\,\mathrm{sign}(w)$ if $0<w^2-\varepsilon^2<4D^2$ and $\eta=1$ otherwise.
For the balanced system $\varepsilon=\epsilon\Li=\epsilon\K$, $w=\omega$, and
$\alpha=1$. The imbalance thus merely scales the Green's function without
causing unexpected features.

The density of states (DOS)
	\begin{align}\label{hpsgen1}
		\nonumber
		h\Li(\omega)&=N\Li\int_{-\infty}^{\infty}d\epsilon\Li\,
		(-1/\pi)\,\mathrm{Im}G^R\Li(k)\\
		&=\frac{N\Li}{\pi}\begin{cases}
			\displaystyle
			2|x|E\left(\frac{x^2}{4}\right) & |x|<2 \\[3mm]
			\displaystyle
			x^2E\left(\frac4{x^2}\right)-(x^2-4)K\left(\frac4{x^2}\right) & |x|>2
		\end{cases}
	\end{align}
is proportional to the DOS of non-interacting lithium $N\Li=m\Li
k_{\mathrm{F}}/(2\pi^2\hbar^2)$ and elliptic functions\footnote{Complete
elliptic functions in the notation of Wolfram Mathematica read
$K\equiv\texttt{EllipticK}$ and $E\equiv\texttt{EllipticE}$.} which depend on
the scaled energy $x=w/D$. As one can see in Fig.~\ref{fig4} (thick line), the
pseudogap has a V shape with states relocated into smooth shoulders with
inflection points at $\omega/D=2$. The gap opens at
$\omega=(m\K\mu\K-m\Li\mu\Li)/(m\K+m\Li)$ and has half-width $2D$. For
comparison, the KM theory results in a pseudogap of the same U shape
\cite{MJL99,CSTL05} as the BCS gap with the half-width $D$.

The walls of the pseudogap \eqref{hpsgen1} are convex functions, in contrast to
the smeared BCS gap in which the concave region extends over more than a half of
the gap bottom. The pseudogap found numerically by Hanai and Ohashi \cite{HO14},
see the lithium DOS in their Fig.~5, has such convex walls. A wider but still
narrow convex region resulted from numerical study of the pure two-dimensional
Li gas, see the low temperature DOS in Fig.~1 of \cite{BPE14}.

The perturbation of the real part of the local Green's function
$g\Li(\omega)=N\Li\int_{-\infty}^{\infty}d\epsilon\Li\,\mathrm{Re}\left[G^R\Li(k
)-G^R\OLi(k)\right]$ is also analytic:
	\begin{multline}\label{hpsgen2}
		g\Li(\omega)=N\Li\theta(2-|x|)\mathrm{sign}(x) \\ \times
		\left[x^2E\left(1-\frac{4}{x^2}\right)-4K\left(1-\frac{4}{x^2}\right)\right].
	\end{multline}
Since the perturbation of the local Green's function
$g^R\Li(\omega)=g\Li(\omega)-i\pi\left[h\Li(\omega)-h\OLi(\omega)\right]$ is
large only in a very narrow region $|\omega|\sim 4D/(\alpha+1/\alpha)$, the
Kramers-Kronig relation $g^R\Li(\omega)=i/\pi\int
d\omega'g^R\Li(\omega')/(\omega-\omega')$ provides an effective way to include
the above neglected finite lifetime due to binary collisions of atoms, by
shifting energies into the complex plane $\omega\to\omega+i/\tau$. We do not
discuss this effect keeping $1/\tau\to 0$.

The T-matrix approach can be modified to cover the superconducting state in
three ways, which we list in order of increasing level of selfconsistency.
First, building the loop in the selfenergy from bare Green's functions one
obtains the KM theory in the version of Patton \cite{Patton71t}. As shown by
Hanai and Ohashi \cite{HO14}, partly selfconsistent theories fail because the
bare Green's function depends on the selfconsistent chemical potential and an
effective Fermi surface of bare potassium atoms is strongly reduced. We remove
this problem enforcing fixed Fermi momentum and show the KM theory for
comparison. The second possibility is to postulate the pseudogap state as the
normal state and add the BCS gap as $\tilde{G}^R\Li(k)=G^R\Li(k)-G^R\Li(k)
\Delta^2G^A\K(-k)\tilde{G}^R\Li(k)$. This renormalized Gor'kov approximation
neglects an effect of the condensate on the non-condensed pairs. The functions
$G^R\Li$ and $G^A\K$ are known \eqref{GpseudogapBal} so that the function
$\tilde{G}^R\Li$ is trivial. In the third approach one leaves Feynman diagrams
and formulate the T-matrix in the multiple scattering expansion of Watson
\cite{GW64}. The multiple scattering corrections vanish in the normal state but
are important for the condensate of Cooper pairs, enabling formation of the BCS
gap \cite{Lipavsky08} which enters the Green's function as
	\begin{multline}\label{GKGordifQ}
		\hat{G}^R\Li(k)=G^R\OLi(k)\\
		-G^R\OLi(k)\left[D^2\hat{G}^A\K(-k)+\Delta^2G^A\K(-k)\right]\hat{G}^R\Li(k).
	\end{multline}
Here hats denote functions with the BCS gap included. Note that the gap does not
enter the potassium propagator in the BCS selfenergy $-\Delta^2G^A\K(-k)$ while
the full selfconsistency appears in the pseudogap selfenergy
$-D^2\hat{G}^A\K(-k)$.

The three above theories can be classified as approximations of the T-matrix
theory in terms of the Nambu-Gor'kov matrices \cite{HPZ09,THO16}. Let us list
them from the best to the crudest. In the Nambu-Gor'kov approach the gap
function depends on energy and momentum due to the non-BCS off-diagonal
selfenergy emerging from relation \eqref{selfenergy1} with the anomalous
function closing the loop. The multiple scattering theory~\eqref{GKGordifQ}
obtains by neglecting this non-BCS off-diagonal term. The Gor'kov and KM
theories obtain by approximating both potassium functions in
Eq.~\eqref{GKGordifQ} by $G^A\K$ and $G^A\OK$, respectively.

\begin{figure}[tb]
\centering{\includegraphics[width=0.8\columnwidth]{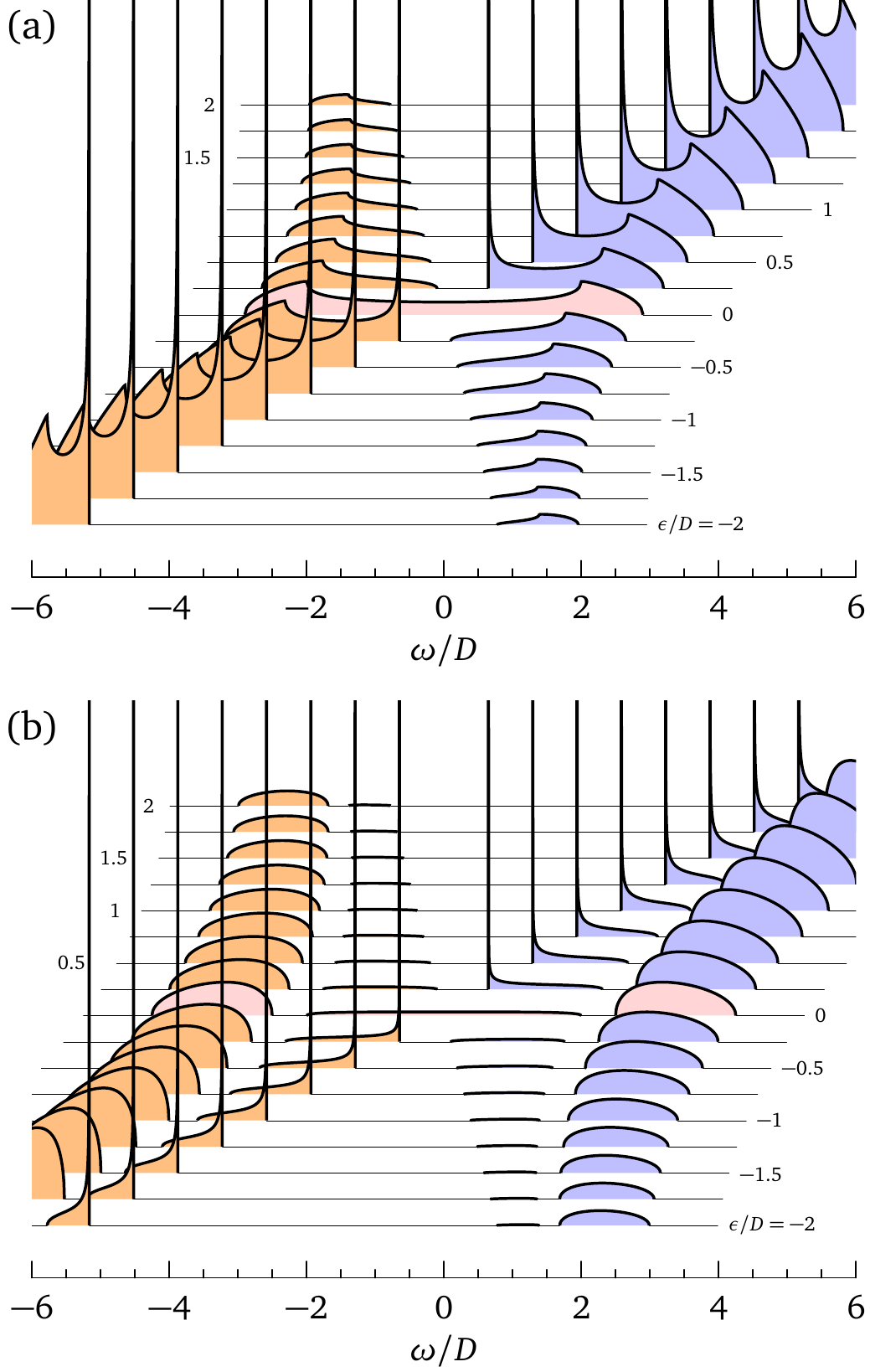}}
\caption{\label{fig3}
Spectral function $\hat{A}\Li(k)$ in the superconducting state for (a)
$\Delta=1.5D$ and (b) $\Delta=3D$, with $\epsilon\Li=\epsilon\alpha$,
$\epsilon\K=\epsilon/\alpha$, and different values of $\epsilon$. The curves are
shifted vertically for clarity. The sharp edges of the pseudogap spectrum
persist in the superconducting state being more suppressed for the larger BCS
gap $\Delta$. For $\Delta>2D$ the spectrum is non-zero in four energy intervals,
for $\Delta<2D$ these intervals merge in two intervals.
}
\end{figure}

\begin{figure}[tb]
\includegraphics[width=0.8\columnwidth]{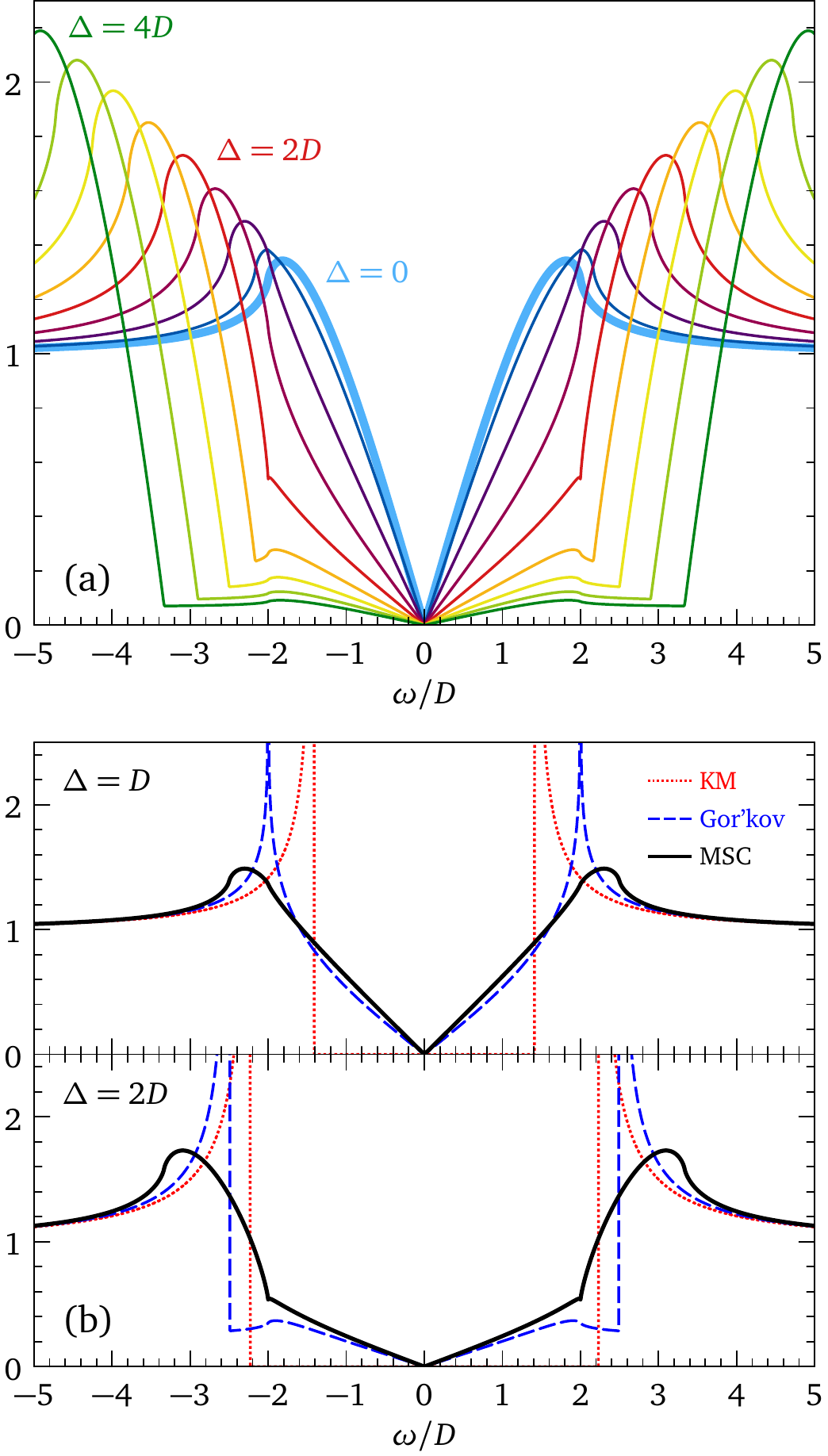}
\caption{\label{fig4}
(a) Density of states for various BCS gaps $\Delta$ in units of the pseudogap
parameter $D$. The `normal'-state pseudogap (thick line) has a V shape. As the
density of the BCS condensate increases it transforms into the U shape. Residual
states in gap have non-analytic features like slope breaks and shoulders of
in-gap states. (b) T-matrix with multiple scattering corrections (MSC) compared
with the Kadanoff-Martin (KM) and Gor'kov theories. In all three theories the
`normal'-state T-matrix is in the static approximation with the same value of
the parameter $D$.
}
\end{figure}

Equation \eqref{GKGordifQ} together with the conjugated equation for
$\hat{G}^A\K(-k)$ are solved by
	\begin{align}
		\label{GFBCSpseudo4}
		\hat{G}^R\Li(k)&=\frac{\omega+\epsilon\K}{2D^2/\Lambda}\left(1-
		\hat{\eta}\sqrt{
		\left|1-\frac{4D^2/\Lambda^2}{(\omega+\epsilon\K)(\omega-\epsilon\Li)}\right|}\,\right), \\
		\nonumber
		\Lambda&=1-\frac{\Delta^2}{2D^2}+\frac{\Delta^2}{2D^2}\eta\sqrt{
		\left|1-\frac{4D^2}{(\omega+\epsilon\K)(\omega-\epsilon\Li)}\right|}\,,\\
		\nonumber
		\hat{\eta}&=\begin{cases}
		e^{\frac{i}{2}\mathrm{arg}
		\left(1-\frac{4D^2/\Lambda^2}{w^2-\varepsilon^2}\right)}&
		0<w^2-\varepsilon^2<4D^2 \\
		i\,\mathrm{sign}(w\Lambda) &
		4D^2<w^2-\varepsilon^2<4D^2/\Lambda^2 \\
		1 & \mathrm{otherwise},
		\end{cases}
	\end{align}
with $\mathrm{arg}(\cdots)\in[-\pi,\pi]$ and $w$, $\varepsilon$ defined before
\eqref{GpseudogapBal}. The spectral function
$\hat{A}\Li(k)=(-1/\pi)\mathrm{Im}\hat{G}^R\Li(k)$ is shown in Fig.~\ref{fig3}
for two values of the BCS gap. The singular edges of the background pseudogap
spectrum are maintained although their amplitude decreases with the increasing
BCS gap. These pseudogap edges result in the slope breaks and in-gap states seen
in the DOS in Fig.~\ref{fig4}. When compared with single-particle spectroscopies
\cite{Torma16}, the sharp edges have to be smeared by the finite lifetime due to
dissipative binary collisions omitted in the present approximation, and by
thermal as well as other extrinsic effects affecting the energy resolution.

The DOS in the superconducting state displayed in Fig.~\ref{fig4}(a) for fixed
pseudogap $D$ and a set of values of the BCS gap $\Delta$ shows that the gap
maintains its V shape for small $\Delta< 2D$, develops into the U-shape gap with
V-shape in-gap states for $\Delta>2D$ which slowly vanish for $\Delta\gg 2D$.
For $\Delta>2D$ our model recalls the DOS of the 2D Hubbard model obtained by a
Monte Carlo calculation restricted to static ($\Omega=0$) correlations, see
Fig.~2(a) in \cite{KM16}. In Fig.~\ref{fig4}(b) one can see that the gap region
is very sensitive to selfconsistency. The DOS resulting from KM theory have the
fixed U shape with effective gap magnitude $\sqrt{\Delta^2+D^2}$. The
renormalized Gor'kov theory shows that the in-gap states are only slightly
modified by the BCS gap effect on the non-condensed pairs, while the gap edges
sharply differ.

In summary, we have derived analytic pseudogap Green's functions for the
`normal' and superconducting states. We have shown that the full selfconsistency
in the closing loop of the selfenergy has pronounced effects on the
single-particle spectrum. The `normal' state is not a Fermi liquid, because the
quasiparticle formation time is the same as the quasiparticle lifetime. The
pseudogap has a V shape in spite of the s-wave pairing. These features contrast
with the partly selfconsistent and partly non-selfconsistent KM theory which
predicts infinitely long living quasiparticles and the U-shaped pseudogap. We
have found that the V-shaped pseudogap has non-trivial interplay with the
U-shaped BCS gap. In particular, the pseudogap features persist in the BCS gap
causing slope breaks of the gap walls and in-gap states. The single-particle
excitations in the superconducting state have non-exponential decay in time so
that they escape a simple approximation by Bogliubov-Valutin quasiparticles.

\section*{Acknowledgments}

The authors are grateful to Madhuparna Karmakar and Pinaki Majumdar for fruitful
comments and to COST action MP1201 and related MSMT-COST LD15062 for financial
support.

\end{document}